%% file: icml25.tex
\documentclass{article} 

\usepackage{microtype}
\usepackage{graphicx}
\usepackage{subcaption}
\usepackage{booktabs} 

\usepackage{hyperref}

\usepackage[accepted]{icml2025} 

\usepackage{amsmath}
\usepackage{amssymb}
\usepackage{mathtools}
\usepackage{amsthm}

\usepackage[capitalize,noabbrev]{cleveref}

\theoremstyle{plain}

\theoremstyle{definition}

\theoremstyle{remark}

\input{math_commands.tex} 

\usepackage[textsize=tiny]{todonotes}

\icmltitlerunning{Dual-Conditional Deep Generation for NIDS Balancing} 

\begin{document}

\twocolumn[
\icmltitle{$\text{C}^{2}\text{BNVAE}$: Dual-Conditional Deep Generation of Network Traffic Data for Network Intrusion Detection System Balancing}

\begin{icmlauthorlist}
\icmlauthor{Yifan Zeng}{sysu}
\end{icmlauthorlist}

\icmlaffiliation{sysu}{Sun Yat-sen University, Guangzhou, China} 

\icmlcorrespondingauthor{Yifan Zeng}{{yifanzeng0615@foxmail.com}} 

\icmlkeywords{Network Intrusion Detection, NIDS, Data Imbalance, Generative Models, VAE, CVAE, Conditional Batch Normalization, Deep Learning}

\vskip 0.3in
]

\printAffiliationsAndNotice{}

\begin{abstract}
Network Intrusion Detection Systems (NIDS) face challenges due to class imbalance, affecting their ability to detect novel and rare attacks. This paper proposes a Dual-Conditional Batch Normalization Variational Autoencoder ($\text{C}^{2}\text{BNVAE}$) for generating balanced and labeled network traffic data. $\text{C}^{2}\text{BNVAE}$ improves the model's adaptability to different data categories and generates realistic category-specific data by incorporating Conditional Batch Normalization (CBN) into the Conditional Variational Autoencoder (CVAE). Experiments on the NSL-KDD dataset show the potential of $\text{C}^{2}\text{BNVAE}$ in addressing imbalance and improving NIDS performance with lower computational overhead compared to some baselines.
\end{abstract}

\section{Introduction}
\label{sec:introduction}
Network intrusions are growing exponentially, diversifying, and becoming more severe. They threaten personal privacy, data security, and can lead to significant economic and social impacts. Deep Learning-based Network Intrusion Detection Systems (DLNIDS) have shown powerful intrusion detection capabilities. However, DLNIDS face a notable challenge in practical applications: the severe class imbalance in network traffic data. In real-world network environments, normal traffic typically constitutes the vast majority, while various types of attack traffic are relatively rare. This imbalance makes it difficult for DLNIDS to effectively learn the features of minority classes, thereby reducing the system's ability to detect novel and rare attacks.

To address data imbalance, researchers have proposed various methods, including resampling techniques (e.g., SMOTE \cite{SMOTE}) and generative approaches \cite{NIDSCVAE,CSAGC}. Among generative methods, Variational Autoencoders (VAEs) \cite{VAE} have shown the ability to learn the latent distribution of data and generate new, synthetic samples. Conditional Variational Autoencoders (CVAEs) \cite{CVAE} further enhance this capability by allowing the generation of data for specific categories using conditional information.

In this paper, we propose $\text{C}^{2}\text{BNVAE}$ (Dual-Conditional Batch Normalization Variational Autoencoder), a generative model designed to effectively produce balanced and labeled network traffic data for DLNIDS. Our primary contribution is the integration of Conditional Batch Normalization (CBN) \cite{CBN} into the CVAE architecture for network traffic data generation. CBN, by learning separate affine transformation parameters ($\gamma, \beta$) for each data category, allows the model to better adapt its normalization process to the specific characteristics of different traffic classes. This "dual-conditional" approach (conditioning in CVAE and conditioning in CBN) aims to enhance the model's adaptability and the realism of generated category-specific data. Through experiments on the NSL-KDD dataset, we validate the potential of $\text{C}^{2}\text{BNVAE}$ in generating minority class samples and subsequently improving NIDS detection performance when using a Decision Tree classifier.

\begin{figure*}[htbp]
    \centering
    \begin{subfigure}[b]{0.49\linewidth}
        \includegraphics[width=\linewidth]{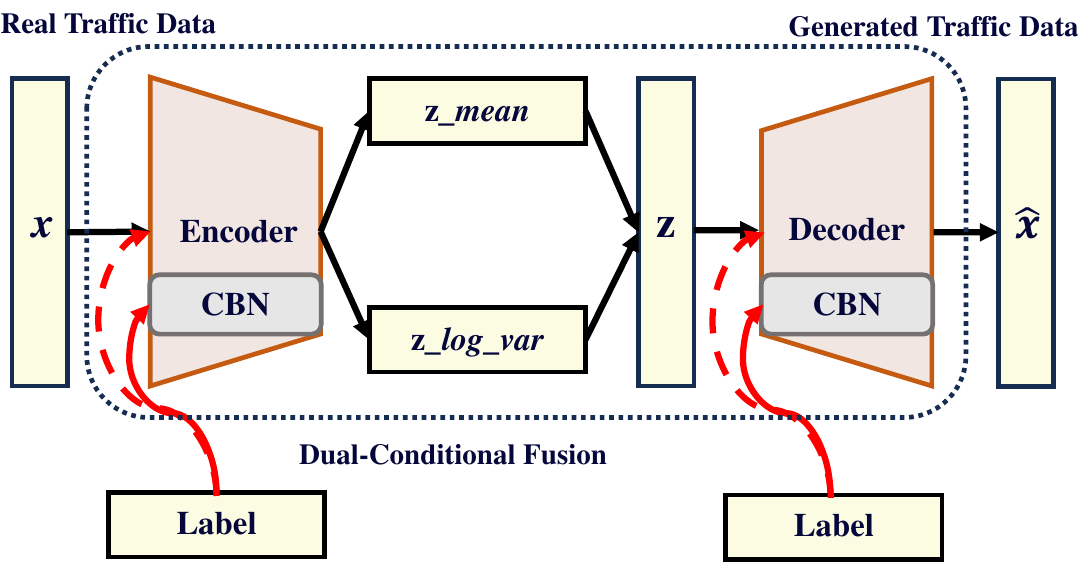} 
        \caption{Overall $\text{C}^{2}\text{BNVAE}$ structure.}
        \label{fig:main_c2bnvae_arch}
    \end{subfigure}
    \hfill 
    \begin{subfigure}[b]{0.47\linewidth}
        \includegraphics[width=\linewidth]{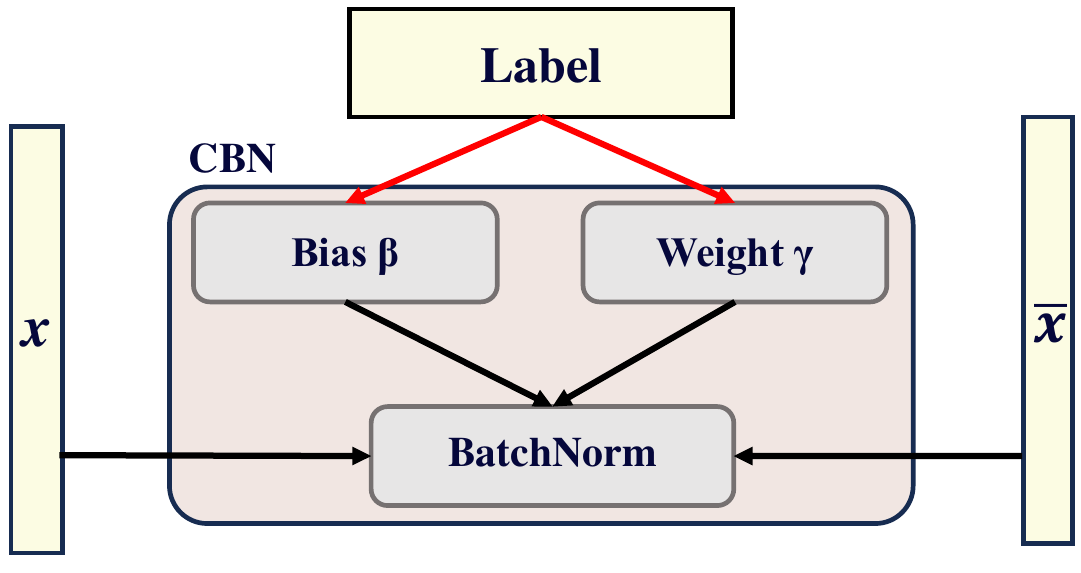} 
        \caption{CBN mechanism.}
        \label{fig:main_cbn_mech}
    \end{subfigure}
    \caption{(a) The overall structure of $\text{C}^{2}\text{BNVAE}$. Real data is input, and generated data is output. Traffic labels are integrated into the Encoder, Decoder, and CBN layers. (b) The structure of CBN. The class label selects the learned scaling factors $\gamma_i$ and $\beta_i$ for normalizing data belonging to that specific class.}
    \label{fig:c2bnvae_structure}
\end{figure*}

\section{Proposed Method: $\text{C}^{2}\text{BNVAE}$}
\label{sec:proposed_method}
The $\text{C}^{2}\text{BNVAE}$ model builds upon the Conditional Variational Autoencoder (CVAE) by incorporating Conditional Batch Normalization (CBN) to enhance category-specific data generation.

\subsection{Conditional Variational Autoencoder (CVAE) Backbone}
A standard VAE learns a probabilistic mapping from the input data $x$ to a latent space $z$ and back. It is trained by maximizing an evidence lower bound (ELBO), which typically consists of a reconstruction term and a regularization term (KL divergence between the learned latent distribution and a prior, usually a standard normal distribution). However, a VAE cannot generate data for specific categories by default.

CVAE extends VAE by conditioning the generation process on additional information, typically class labels $y$. In our $\text{C}^{2}\text{BNVAE}$, the one-hot encoded class label $y$ of the traffic data is concatenated with the input $x$ for the encoder and with the latent variable $z$ for the decoder. This allows the model to learn class-specific latent representations and generate samples belonging to a target minority class by providing its label $y$.
The loss function for $\text{C}^{2}\text{BNVAE}$:
\begin{equation}
    \mathcal{L}_{\text{total}} = \mathcal{L}_{\text{recon}} + \mathcal{L}_{\text{regu}}
    \label{eq:total_loss}
\end{equation}
where $\mathcal{L}_{\text{recon}}$ is the reconstruction loss and $\mathcal{L}_{\text{regu}}$ is the regularization loss.
Let $N$ be the number of training samples, $x_i$ be the $i$-th training sample, and $\hat{x}_i$ be the $i$-th sample generated by the model conditioned on $y_i$. $\mathcal{L}_{\text{recon}}$ is measured by the mean squared error:
\begin{equation}
    \mathcal{L}_{\text{recon}} = \frac{1}{N} \sum_{i=1}^{N} (x_i - \hat{x}_i)^2
\end{equation}
$\mathcal{L}_{\text{regu}}$ is the KL divergence between the learned posterior distribution of latent variables $q(z|x,y)$ (approximated by $N(\mu(x,y), \sigma^2(x,y))$) and the prior $p(z|y)$ (often simplified to $N(0, I)$):
\begin{equation}
    \mathcal{L}_{\text{regu}} = \mathcal{D}_\text{KL}(q(z|x,y) \mid\mid p(z|y))
\end{equation}

\subsection{Conditional Batch Normalization (CBN)}
In standard CVAEs, Batch Normalization (BN) \cite{BN} is often used to stabilize training and improve generalization. However, BN applies a single set of learned affine parameters ($\gamma, \beta$) across all samples in a batch, regardless of their class. This can be suboptimal when generating data for diverse categories, potentially leading to more uniform or less distinct category-specific features.

CBN addresses this by making the affine transformation parameters conditional on the class label $y$. For each class $i$, CBN learns separate scaling ($\gamma_i$) and shifting ($\beta_i$) parameters. Given a mini-batch input $x$, its sample mean $\hat{\boldsymbol{\mu}}$ and standard deviation $\hat{\boldsymbol{\sigma}}$, the CBN transformation for samples belonging to class $i$ is:
\begin{equation}
   \overline{x} = \mathrm{CBN}({x}|y=i) = \gamma_{i} \frac{{x} - \hat{\boldsymbol{\mu}}}{\sqrt{\hat{\boldsymbol{\sigma}}^2 + \epsilon}} + \beta_{i}
   \label{eq:cbn_formula_main}
\end{equation}
where $\epsilon$ is a small constant for numerical stability. By incorporating CBN into the decoder (and potentially encoder) layers of the CVAE, $\text{C}^{2}\text{BNVAE}$ aims to provide greater flexibility in modeling the distinct statistical properties of each traffic category. This "dual conditioning" – first through the CVAE structure and second through CBN – is hypothesized to improve the model's adaptability and the diversity and authenticity of the generated samples, especially for minority classes.

The overall structure of $\text{C}^{2}\text{BNVAE}$ is depicted in \cref{fig:c2bnvae_structure}.

$\text{C}^{2}\text{BNVAE}$ aims to:
\begin{itemize}
    \item Generate high-quality, realistic traffic data for specific (minority) categories.
    \item Balance imbalanced datasets by synthesizing minority class samples, thereby enhancing the detection performance of downstream NIDS classifiers.
    \item Create more diverse and authentic synthetic data that accurately reflects the features and categorical distinctions within network traffic.
\end{itemize}

\section{Experimental Setup}
\label{sec:experimental_setup}

\subsection{Dataset}
We conducted experiments on the widely used NIDS benchmark dataset, NSL-KDD \cite{NSL-KDD}. NSL-KDD is an improved version of the KDD Cup '99 dataset and is frequently used for evaluating intrusion detection systems. It contains various types of attacks and normal traffic records, characterized by 41 features. A key characteristic of NSL-KDD is its significant class imbalance, making it a suitable benchmark for assessing algorithms designed to handle imbalanced data. For our experiments, we used the KDDTrain+ dataset for training the generative models and the KDDTest+ for evaluating the NIDS classifier. To balance the dataset for training the classifier, with the Normal class in KDDTrain+ having the most samples (67343), we generated a matching number of samples for each minority attack category using the respective data augmentation algorithms.

\subsection{Evaluation Metrics}
To evaluate the performance of the NIDS classifier trained on data augmented by different methods, we employed comprehensive metrics suitable for imbalanced datasets:
\begin{itemize}
    \item \textbf{Accuracy (Acc):} The proportion of correctly classified instances among the total instances:
    \begin{equation}
        \text{Acc} = \frac{\text{TP} + \text{TN}}{\text{TP} + \text{TN} + \text{FP} + \text{FN}}
    \end{equation}
    
    \item \textbf{Weighted Precision ($\text{Pre}_{w}$):} The weighted average of precision for each class, where Pre is TP / (TP + FP):

     \begin{equation}
        \text{Pre}_w = \sum_{c=1}^{C} \left( \frac{N_c}{N_{\text{total}}} \cdot \frac{\text{TP}_c}{\text{TP}_c + \text{FP}_c} \right)
    \end{equation}
    
    \item \textbf{Weighted Recall ($\text{Recall}_{w}$):} The weighted average of recall for each class, where Recall is TP / (TP + FN):
    \begin{equation}
        \text{Recall}_w = \sum_{c=1}^{C} \left( \frac{N_c}{N_{\text{total}}} \cdot \frac{\text{TP}_c}{\text{TP}_c + \text{FN}_c} \right)
    \end{equation}
    
    \item \textbf{Weighted F1-Score ($\text{F1}_{w}$):} The weighted average of the F1-score for each class. The F1-score is the harmonic mean of precision and recall ($2 \times (Pre \times Recall) / (Pre + Recall)$). The weighted F1-score is calculated as:
    \begin{equation}
        \text{F1}_{w} = \sum_{c=1}^{C} \left( \frac{N_c}{N_{\text{total}}} \cdot 2 \cdot \frac{\text{Pre}_c \cdot \text{Recall}_c}{\text{Pre}_c + \text{Recall}_c} \right)
        \label{eq:f1_weighted_main}
    \end{equation}
    where $C$ is the number of classes, $N_c$ is the number of instances in class $c$, and $N_{\text{total}}$ is the total number of instances.

    \item \textbf{FLOPs:} Floating-point operations, measuring computational complexity during inference. Lower values indicate higher efficiency.
    \item \textbf{Params:} Number of trainable parameters, reflecting model size and memory requirements.
\end{itemize}
These weighted metrics give more importance to classes with more samples but still provide a holistic view of performance across all classes.

\subsection{Baselines and Classifier}
We compared $\text{C}^{2}\text{BNVAE}$ with several baseline data balancing techniques: Original imbalanced data (No balancing), Random Oversampling, SMOTE (Synthetic Minority Over-sampling Technique) \cite{SMOTE}, Borderline SMOTE \cite{Borderline}, KMeans SMOTE \cite{Kmeans}, SVM SMOTE \cite{SVMSMOTE}, Standard CVAE (without CBN) \cite{CVAE}.

The downstream NIDS classifier employed in all experiments is a Decision Tree (DT). We chose DT for its simplicity, interpretability, and common use as a baseline classifier. The performance of the DT classifier trained on data generated/balanced by these algorithms serves as a proxy for the quality and utility of the augmented data.

\subsection{Implementation Details}
The experiments were conducted on a computing environment with an Intel (R) Xeon (R) Gold 6240 CPU @ 2.60GHz and a Tesla V100S-PCIE-32GB GPU. The operating system was Ubuntu 18.04.3 LTS. All code was implemented in Python 3.7.6 using PyTorch 1.13.1+cu117. Specific model hyperparameters for $\text{C}^{2}\text{BNVAE}$ are detailed in \cref{tab:hyperparameters_main}.


\begin{table}[t]
\centering
\caption{Model Hyperparameters of $\text{C}^{2}\text{BNVAE}$}
\label{tab:hyperparameters_main}
\begin{tabular}{lcc}
\toprule
\bf Hyperparameters & \bf Encoder & \bf Decoder \\
\midrule
Layers & [128,60$\times$4,32] & [37,60$\times$4,123] \\
Activation & LeakyReLU & LeakyReLU \\
Initialization & He & He \\
Batch size & 128 & 128 \\
Learning rate & 1e-4 & 1e-4 \\
Epoch & 120 & 120 \\
Optimizer & Adam & Adam \\
Loss function & MSELoss & MSELoss \\
\bottomrule
\end{tabular}
\end{table}

\section{Results and Discussion}
\label{sec:results_discussion}

\subsection{Computational Overhead}
$\text{C}^{2}\text{BNVAE}$ has a total of 43,627 parameters and incurs 43,200 FLOPs per sample. The encoder component contributes 22,744 parameters and 22,560 FLOPs, while the decoder has 20,883 parameters and 20,640 FLOPs. For comparison, a Conditional Generative Adversarial Network (CGAN) baseline \cite{CGAN,CSAGC}, implemented with a generator having layers [128, 100$\times$5, 123] and a discriminator with layers [128, 100, 50, 50, 1], has significantly more parameters (87,820) and higher computational cost (109,892 FLOPs). This comparison highlights the relatively lower computational overhead of the proposed $\text{C}^{2}\text{BNVAE}$ model, making it potentially more efficient for deployment in resource-constrained NIDS environments.

\subsection{Intrusion Detection Performance}
The primary evaluation of $\text{C}^{2}\text{BNVAE}$ lies in its ability to generate synthetic minority class data that, when used to balance the training set, improves the performance of a downstream NIDS classifier. \cref{tab:main_results} presents the intrusion detection performance of the Decision Tree classifier on the KDDTest+ dataset when trained on data augmented by various methods.

\begin{table}[t]
\centering
\caption{Intrusion detection performance experimental results (\%) on KDDTest+ using a Decision Tree classifier trained on augmented KDDTrain+ data.}
\vspace{10pt}
\label{tab:main_results}
\resizebox{\columnwidth}{!}{
\begin{tabular}{l c c c c}
\toprule
\bf Algorithms & \bf Acc & \bf $\text{Pre}_{w}$ & \bf $\text{Recall}_{w}$ & \bf $\text{F1}_{w}$ \\ 
\midrule
Original imbalanced Data & 75.88 & 79.32 & 75.88 & 72.74 \\
Random oversampling & 77.02 & 79.14 & 77.02 & 73.84 \\
SMOTE & 76.11 & 78.04 & 76.11 & 73.33 \\
Borderline SMOTE & 75.89 & 78.73 & 75.89 & 73.59 \\
KMeans SMOTE & 76.13 & 79.03 & 76.13 & 72.81 \\
SVM SMOTE & 78.11 & 79.09 & 78.11 & 76.02 \\
CVAE & 78.45 & 79.10 & 78.45 & 77.18 \\
\bf $\text{C}^{2}\text{BNVAE}$ & \bf 79.40 & \bf 80.69 & \bf 79.40 & \bf 78.19 \\
\bottomrule
\end{tabular}
} 
\end{table}

As shown in \cref{tab:main_results}, training the Decision Tree classifier on data augmented by $\text{C}^{2}\text{BNVAE}$ yielded the best performance across all evaluated metrics: Accuracy (79.40\%), Weighted Precision (80.69\%), Weighted Recall (79.40\%), and Weighted F1-Score (78.19\%). This represents a notable improvement over training on the original imbalanced data (F1-Score: 72.74\%) and also surpasses other common oversampling techniques like Random Oversampling (F1-Score: 73.84\%) and various SMOTE variants (F1-Scores ranging from 72.81\% to 76.02\%).

Importantly, $\text{C}^{2}\text{BNVAE}$ also outperforms the standard CVAE (F1-Score: 77.18\%). This suggests that the introduction of Conditional Batch Normalization (CBN) provides a tangible benefit in generating more effective synthetic samples for balancing the dataset. The CBN allows the model to learn class-specific normalization parameters, potentially leading to the generation of minority class samples that are more distinct and better capture the unique characteristics of each attack type. This, in turn, helps the downstream classifier to learn more robust decision boundaries, particularly for underrepresented attack classes.

\begin{figure}[t]
    \centering
    \includegraphics[width=\columnwidth]{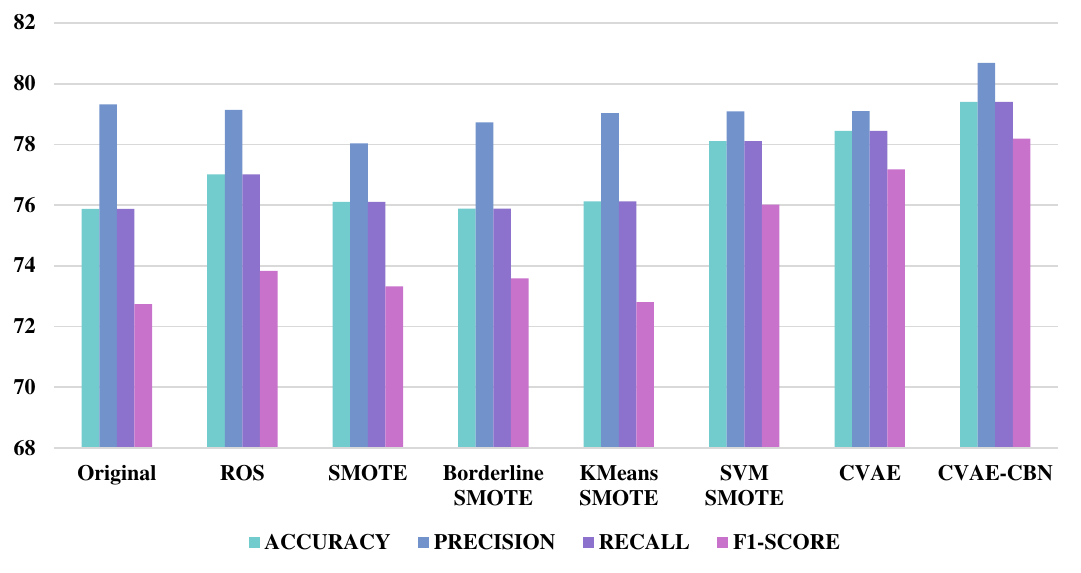} 
    \caption{Bar chart of Decision Tree detection performance when trained on balanced data processed by different algorithms. Higher bars indicate better performance.}
    \label{fig:results_bar_chart_main}
\end{figure}

The bar chart in \cref{fig:results_bar_chart_main} further visualizes these performance differences, clearly positioning $\text{C}^{2}\text{BNVAE}$ as the top-performing data augmentation method in this experimental setup.

\subsection{Discussion}
The superior performance of $\text{C}^{2}\text{BNVAE}$ can be attributed to its dual-conditional mechanism. The CVAE component ensures that generated samples belong to the specified target class, while the CBN component fine-tunes the feature generation process by applying class-specific normalization. This likely results in synthetic data that not only increases the representation of minority classes but also maintains or even enhances the separability between different classes. Traditional BN, used in standard CVAE, might inadvertently smooth out some class-specific features by applying uniform normalization parameters. CBN mitigates this by allowing the normalization to adapt to each class's statistical profile, leading to more realistic and useful synthetic samples.

The lower computational overhead of $\text{C}^{2}\text{BNVAE}$ compared to GAN \cite{GAN} or CGAN is also a practical advantage, particularly for NIDS applications where frequent retraining or deployment on edge devices might be necessary.

While these results are promising, it is important to acknowledge that they are based on the NSL-KDD dataset and a Decision Tree classifier. The effectiveness of $\text{C}^{2}\text{BNVAE}$ might vary with other datasets possessing different characteristics or when paired with more complex deep learning classifiers for NIDS. Future work should explore its generalizability across a wider range of network environments, attack types, and NIDS architectures. Additionally, qualitative analysis of the generated samples (e.g., using t-SNE or UMAP for visualization) could provide further insights into how well $\text{C}^{2}\text{BNVAE}$ captures the underlying data distributions compared to other methods.

\section{Conclusion}
\label{sec:conclusion}
In this paper, we proposed $\text{C}^{2}\text{BNVAE}$, a dual-conditional generative model for addressing class imbalance in network traffic data for NIDS. By integrating CBN into a CVAE, $\text{C}^{2}\text{BNVAE}$ enhances the model's adaptability to different data categories and its ability to generate realistic, category-specific synthetic samples. Our experimental results on the NSL-KDD dataset demonstrate that $\text{C}^{2}\text{BNVAE}$ outperforms several baseline oversampling techniques and standard CVAE in improving the detection performance of a Decision Tree based NIDS, while also exhibiting lower computational overhead compared to a CGAN baseline. This study highlights the potential of incorporating class-conditional normalization techniques like CBN into generative models for effectively tackling data imbalance in security applications. While further research is needed to assess its generalizability, $\text{C}^{2}\text{BNVAE}$ offers a promising and computationally efficient approach for enhancing the robustness of NIDS against novel and rare attacks by creating more balanced and representative training datasets.


\bibliography{example_paper} 
\bibliographystyle{icml2025}




\end{document}

%% file: math_commands.tex
\usepackage{amsmath,amsfonts,bm}

\def\eqref#1{equation~\ref{#1}}

\def\1{\bm{1}}

\DeclareMathAlphabet{\mathsfit}{\encodingdefault}{\sfdefault}{m}{sl}
\SetMathAlphabet{\mathsfit}{bold}{\encodingdefault}{\sfdefault}{bx}{n}

